\documentclass[twocolumn,10pt]{asme2e}

\usepackage{booktabs,amsmath,amssymb,amsbsy,latexsym,graphicx,color}
\usepackage{rotating}
\usepackage{float}
\usepackage{subfigure}
\usepackage{bm}
\usepackage{upgreek}
\usepackage{pbox}
\usepackage{newfloat}
\usepackage{nomencl}
\usepackage{indentfirst}
\usepackage{etoolbox}
\usepackage{tocvsec2}
\usepackage[titletoc]{appendix}
\usepackage{appendix}
\usepackage[bbgreekl]{mathbbol}

\confshortname{SMASIS 2018}
\conffullname{the ASME 2018 Conference on Smart Materials, Adaptive Structures and Intelligent Systems}

\confdate{10-12}
\confmonth{September}
\confyear{2018}
\confcity{San Antonio, Texas}
\confcountry{USA}

\papernum{SMASIS2018-8050}

\title{A Three-dimensional Constitutive Model for Polycrystalline Shape Memory Alloys Under Large Strains Combined With Large Rotations}

\author{Lei Xu
    \affiliation{
	Dept. of Aerospace Engineering\\
	Texas A\&M Univeristy\\
	College Station, Texas 77843\\
    Email: sdf007xulei@tamu.edu
    }	
}

\author{Theocharis Baxevanis
	\affiliation{
		Dept. of Mechanical Engineering\\
		University of Houston\\
		Houston, Texas 77204\\
	}	
}

\author{Dimitris Lagoudas
	\affiliation{
		Dept. of Aerospace Engineering\\
		Dept. of Material Science \& Engineering\\
		Texas A\&M Univeristy\\
	    College Station, Texas 77843\\
	}	
}

\begin{document}

\maketitle    

\begin{abstract}
{Shape Memory Alloys (SMAs), known as an intermetallic alloys with the ability to recover its predefined shape under specific thermomechanical loading, has been widely aware of working as actuators for active/smart morphing structures in engineering industry. Because of the high actuation energy density of SMAs, compared to other active materials, structures integrated with SMA-based actuators has  high advantage in terms of trade-offs between overall structure weight, integrity and functionality. The majority of available constitutive models for SMAs are developed within infinitesimal strain regime. However, it was reported that particular SMAs can generate transformation strains nearly up to 8\%-10\%, for which the adopted infinitesimal strain assumption is no longer appropriate. Furthermore, industry applications may require SMA actuators, such as a SMA torque tube, undergo large rotation deformation at work. Combining the above two facts, a constitutive model for SMAs developed on a finite deformation framework is required to predict accurate response for these SMA-based actuators under large deformations.

A three-dimensional constitutive model for SMAs considering large strains with large rotations is proposed in this work. This model utilizes the logarithmic strain as a finite strain measure for large deformation analysis so that its rate form hypo-elastic constitutive relation can be consistently integrated to deliver a free energy based hyper-elastic constitutive relation. The martensitic volume fraction and the second-order transformation strain tensor are chosen as the internal state variables to characterize the inelastic response exhibited by polycrystalline SMAs. Numerical experiments for basic SMA geometries, such as a bar under tension and a torque tube under torsion are performed to test the capabilities of the newly proposed model. The presented formulation and its numerical implementation scheme can be extended in future work for the incorporation of other inelastic phenomenas such as transformation-induced plasticity, viscoplasticity and creep under large deformations.}
\end{abstract}

\section*{NOMENCLATURE}
\begin{tabbing}
	\bfseries left \quad \=\bfseries center \quad \=\bfseries right \quad \=\bfseries paragraph \kill
	$\mathbf{B}$ \> \quad Left Cauchy-Green tensor \\ 
	$\mathbf{B}_{i,j}$ \> \quad Subordinate eigenprojections of $\mathbf{B}$  \\  
	$\mathbf{D}$ \>\quad Rate of deformation tensor \\
	$\mathbf{D}^e$ \>\quad Rate of deformation tensor of elastic part  \\
	$\mathbf{D}^{tr}$ \>\quad Rate of deformation tensor of transformation part \\
	$\mathbf{F}$ \>\quad Deformation gradient \\
	$\mathbf{L}$ \>\quad Velocity gradient\\  
	$\mathcal{S}$ \>\quad Effective compliance tensor\\  
	$\mathcal{S}^A$ \>\quad Compliance tensor for austenite phase\\  
	$\mathcal{S}^M$ \>\quad Compliance tensor for martensite phase\\ 
	$\Delta \mathcal{S}$ \>\quad Compliance tensor phase difference\\
	$\mathbf{W}$ \>\quad Anti-symmetric part of velocity gradient\\
	$\mathbf{X}$ \>\quad Position vector in reference configuration\\
	$\bm\Lambda$ \>\quad Transformation direction tensor\\
	$\bm\Lambda^{fwd}$ \>\quad Forward transformation direction tensor\\
	$\bm\Lambda^{rev}$ \>\quad Reverse transformation direction tensor\\
	$\bm{\Omega}^{log}$ \>\quad Logarithmic spin\\
	$\mathbf{h}$ \>\quad Logarithmic strain of Eulerian type \\
	$\mathbf{h}^{tr}$ \>\quad Transformation strain \\ 
	$\mathbf{h}^{tr-r}$ \>\quad Transformation strain at reverse point \\  
	$\bar{\mathbf{h}}^{tr-r}$ \>\quad Effective transformation strain at reverse point \\      
	$\mathbf{x}$ \>\quad Position vector at current configuration\\

	$\bm{\Upsilon}$ \>\quad Internal state variable set\\
	$\bm{\uptau}$ \>\quad Kirchhoff stress tensor \\ 
	$\bm{\uptau}'$ \>\quad Deviatoric part of Kirchhoff stress \\ 
	${\bar\uptau}^{'}$ \>\quad Effective Kirchhoff stress \\   
	$\bm\alpha^A$ \>\quad Second order thermal expansion for austenite   \\
	$\bm\alpha^M$ \>\quad Second order thermal expansion for martensite  \\  
	$\Delta\alpha$ \>\quad Thermal expansion phase difference\\      
	$\mathcal{D} $ \>\quad Dissipation energy\\   
	
	$A_s$ \>\quad Austenite phase transformation start temperature\\
	$A_f$ \>\quad Austenite phase transformation finish temperature\\  
	$M_s$ \>\quad Martensite phase transformation start temperature\\
	$M_f$ \>\quad Martensite phase transformation finish temperature\\       
	$G$ \>\quad Gibbs free energy\\
	$H^{max}$ \>\quad Maximum transformation strain\\  
	$T$ \>\quad Temperature \\ 
	$T_0$ \>\quad Temperature at reference point \\ 
	$Y$ \>\quad Critical thermodynamic driving force \\   
	
	$a_1,a_2,a_3$ \>\quad Material parameters in hardening function \\ 
	$c$ \>\quad Specific heat \\            
	$f(\xi)$ \>\quad Hardening function \\     
	$s$ \>\quad Specific entropy \\
	$s_0$ \>\quad Specific entropy at reference state \\  
	$\Delta s_0$ \>\quad Specific entropy phase difference \\    
	$u$ \>\quad Internal energy\\
	$u_0$ \>\quad Internal energy at reference state\\  
	$\Delta u_0$ \>\quad Internal energy phase difference \\

	$\Phi$ \>\quad Transformation function\\
	$\rho$ \>\quad Density at current configuration  \\
	$\rho_{0}$ \>\quad Density at reference configuration  \\
	$\xi$ \>\quad Martensite volume fraction  \\
	$\nabla$ \>\quad Gradient operator\\
	$\chi$ \>\quad Deformation mapping function\\
	$\lambda_i,\lambda_j$ \>\quad Eigenvalues of $\mathbf{B}$  \\
	$\pi$ \>\quad Thermodynamic driving force\\  
\end{tabbing}

\section*{INTRODUCTION} \label{sec:intro} 

Shape Memory Alloys (SMAs), known as an intermetallic alloys with the ability to recover its predefined shape under specific thermomechanical loading, has been widely aware of working as actuators for active/smart morphing structures in engineering industry. Because of the high actuation energy density of SMAs, structures integrated with SMA-based actuators has high advantage in terms of trade-offs between overall structure weight, integrity and functionality. There has been existing fatigue, damage\cite{wheeler2014characterization,Darren2018} and preliminary corrosion research\cite{MM2018} into SMAs over the past several decades in order to realize SMA-based actuators into extensive industry applications.

A substantial of SMAs constitutive theories at continuum levels have been proposed, the majority of them are within small deformation regime based on infinitesimal strain assumption. Some thorough review of shape memory models can be found from Boyd and Lagoudas\cite{boyd1996}, Birman and November\cite{birman1997}, Raniecki and Lexcellent \cite{raniecki1992,raniecki1998}, Patoor et al.\cite{patoor2006,patoor1996}, Hackl and Heinen\cite{hackl2008}, Levitas and Preston \cite{levitas1998} etc.. The aforementioned models are able to predict SMAs response accurately within infinitesimal strain regime. However, recent publication has reported that shape memory alloys can reversibly deform to a relatively large strain up to 8\% \cite{jani2014,shaw2000}, and also repeated cycling loading has been reported to induce irrecoverable transformation induced plasticity strains up to 20\% or more\cite{wheeler2014,Xu2017trip}. Strain regime in cracked SMA specimen can also go up to 8\%\cite{BH1,BH2,BH3,BH4}. In addition to such relatively large strains, SMAs-based devices may also undergo large rotations during its deployment. Combining above two factors, it is indispensable to develop a constitutive model based on finite deformation framework to provide accurate predictions for these SMA-based actuators under large deformations. 

Much efforts has been devoted to proposing such a constitutive model for SMAs at finite deformation framework. Following finite deformation theory, two commonly accepted kinematic assumptions are usually adopted for the model development. One is the multiplicative decomposition on deformation gradient, and the other one is the additive decomposition on the total rate of deformation tensor. The first one is simply based on classic crystal-plasticity theory. In contrast, the second one is following energy conservation principle that the total energy provided from the outside can be additively divided into a recoverable part plus a dissipative part.

Existing approaches based on the multiplicative decomposition to formulate a finite strain constitutive model for SMAs can be found from literature Auricchio \cite{auricchio1997,auricchio2001}, Arghavani et al.\cite{arghavani2011,arghavani2011improved,arghavani20103d}, Thamburaja and Anand\cite{tham2010,tham2001}, Reese and Christ\cite{christ2009,reese2008} etc.. However, the complexity of this model type and its high computational cost hinders its wide usage for industry application analysis that involved with a large amount of trial and error computation. In contrast, finite strain constitutive models based on additive composition achieves a much simpler model structure and easier implementation procedures, which enables it gaining a lot popularity and is commonly used in current available commercial finite element analysis softwares, such as Abaqus and ANSYS. However, this model types requires to adopt an objective rate in its rate form hypoelastic constitutive equation to achieve the principle of objectivity. A number of objective rates, such as Zaremba-Jaumann-Noll rate, Green-Naghdi-Dienes rate and Truesdell rate etc., have been proposed to meet this objectivity goal. However, those objective rates aforementioned are not real 'objective' for their failure to integrate the rate form hypo-elastic equation to deliver a free energy based hyperelastic stress-strain relation\cite{xiao2006}. Therefore many spurious phenomenons, such as artificial residual stress accumulation and shear stress oscillation, are often observed even for indissipative elastic material undergoing closed path cyclic loadings.

It was not until the logarithmic rate proposed and the proof conducted by Xiao et al.\cite{xiao1997,xiao1997hypo}, Bruhns et al.\cite{bruhns1999self,bruhns2001large,bruhns2001self} and Meyers et al.\cite{meyers2003elastic,meyers2006choice}, that such self-inconsistency issue related to objective rates has been resolved. In their mathematical proof from publication\cite{xiao1997} in the year of 1997, it was shown that: the logarithmic rate of logarithmic strain of its Eulerian type is exactly identical with the rate of deformation tensor, and logarithmic strain is the only one among all other strain measures enjoying this important property. This new development in finite deformation theory not only provides consistent solutions to classical plasticity problems for metallic material, but also sheds lights on the development of finite strain constitutive model for SMAs. Available yet still limited publications for SMAs model developed along this line can be obtained from M\"uller and Bruhns\cite{muller2006}, Teeriaho\cite{teeriaho2013} and Xu et al. \cite{xu2017}.

In this article, a three-dimensional constitutive model for SMAs considering large strains with large rotations is going to be proposed. This model utilizes the logarithmic strain as a finite strain measure for large deformation analysis so that its rate form hypoelastic constitutive relation can be consistently integrated to deliver a free energy based hyperelastic constitutive relation. The martensitic volume fraction and the second-order transformation strain tensor are chosen as the internal state variables to characterize the inelastic response exhibited by polycrystalline SMAs. Numerical experiments for basic geometries, such as a bar under tension and a torque tube under torsion are performed to test the capabilities of the newly proposed model.

\section*{PRELIMINARY} \label{Preliminary}
\subsection*{Continuum Mechanics}
Let body $ \mathcal{B} $ with its material point defined by a position vector $ \mathbf{X} $ in the reference (undeformed) configuration at time $ t_{0} $, let vector $ \mathbf{x} $ represent the position vector occupied by that material point in current (deformed) configuration at time $ t $, therefore the deformation process of the material point from its initial configuration to current configuration can be defined through the well known second order deformation gradient tensor $\mathbf{F}(\mathbf{x},t)$:
\begin{equation}\label{Deformation}
\mathbf{F}(\mathbf{x},t) =\frac{\partial \mathbf{x}}{ \partial \mathbf{X}}  
\end{equation}
The velocity field of the material point can be defined by the second order tensor $\mathbf{v}$ as,
\begin{equation}\label{Velocity}
\mathbf{v} = \dfrac{d \mathbf x}{d t} = \dot{\mathbf x}
\end{equation}
Based on the velocity tensor $\mathbf{v}$, the velocity gradient $\mathbf{L}$ can be derived as,
\begin{equation}\label{eq:V_gradient}
\mathbf{L}= \frac{\partial \mathbf{v}}{ \partial \mathbf{x}}= \mathbf{\dot{F}}\mathbf{F} ^{-1}  
\end{equation}
The following equation on polar decomposition of deformation gradient is well known,
\begin{equation}\label{eq:PolarDecom}
\mathbf{F  = UR = VR}
\end{equation}
In Eqn.(\ref{eq:PolarDecom}), the second order orthogonal tensor $\mathbf{R}$ is called the rotation tensor, i.e., $\mathbf{R}\mathbf{R^T}=\mathbf{R^T}\mathbf{R}=\mathbf{I}$, where $\mathbf{I} $ is second order identity tensor.  Symmetric and positive definite second order tensors $\mathbf{U}$ and $\mathbf{V}$ are called right (or Lagrangian) and left (or Eulerian) stretch tensors, through which the right Cauchy-Green tensor $\mathbf{C}$ and the left Cauchy-Green tensor $\mathbf{B}$ are obtained,
\begin{equation}\label{RCG}
\mathbf{C} = \mathbf{F^{T}F} =\mathbf{U}^2 
\end{equation}
\begin{equation}\label{LCG}
\mathbf{b} = \mathbf{FF^{T}} =\mathbf{V}^2 
\end{equation}
The logarithmic (or Hencky) strain of its Lagrangian type $ \mathbf{H} $ and Eulerian type $ \mathbf{h} $ are calculated as,
\begin{equation}\label{L_LogStrain}
\mathbf{H} = \frac{1}{2} \ln(\mathbf{ {C}}) =\mathbf{\ln ({U})}
\end{equation}
\begin{equation}\label{E_LogStrain}
\mathbf{h} = \frac{1}{2} \ln(\mathbf{ {b}}) =\mathbf{\ln ({V})}
\end{equation}
It is known velocity gradient $\mathbf{L}$ can be additively decomposed into a symmetric part called the rate of deformation tensor $\mathbf{D}$ plus an anti-symmetric part named spin tensor $\mathbf{W}$. 
\begin{equation}\label{S_P_tensor}
\mathbf{L = D + W}, \qquad
\mathbf{D} =\dfrac{1}{2} \mathbf{(L+L^{T})}, \vspace{5pt}  \qquad 
\mathbf{W} =\dfrac{1}{2} \mathbf{(L-L^{T})}  
\end{equation}

\subsection*{Logarithmic Rate and Logarithmic Spin}\label{subsec:objective_rate}
As it was discussed in introduction, two commonly accepted kinematic assumption are usually considered in finite deformation theory. The first one is based on multiplicative decomposition of deformation gradient $\mathbf{F}$, and the other one is based on additive decomposition of total rate of deformation tensor $\mathbf{D}$. For a long period of time, rate form hypoelastic constitutive theory has been criticized for its failure to be exactly integrated to define an elastic material behavior\cite{simo2006}, this includes many well known objective rates such as Zaremba-Jaumann rate, Green-Naghdi rate and Truesdell rate etc.. Because of the aforementioned issues, many spurious dissipative phenomenons, such as shear stress oscillation and accumulated artificial residual stress are observed in even simple elastic material deformation. In other words, the non-integrable hypoelastic constitutive equation that uses objective rates is path-dependent and dissipative, and thus would deviate essentially from the recoverable elastic-like behavior \cite{xiao2006}.

As it was proved in the work by Xiao et al.\cite{xiao1997,xiao1997hypo,xiao2006}, Bruhns et al.\cite{bruhns1999self,bruhns2001large,bruhns2001self} and Meyers et al.\cite{meyers2003elastic,meyers2006choice}, the logarithmic rate of the logarithmic strain $\mathbf{h}$ of its Eulerian type is identical with the rate of deformation tensor $\mathbf{D}$, by which a grade-zero hypoelastic model can be exactly integrated into an finite deformation elastic model\cite{xiao1997}. This unique relationship between logarithmic strain $\mathbf{h}$ and the rate of deformation tensor $\mathbf{D}$ can be expressed as,
\begin{equation}\label{eq:Log_strain_rate}
\mathring{\mathbf{h}}^{log} = \dot{\mathbf{h}}+\mathbf{h} \bm{ \Omega}^{log}-\bm{ \Omega}^{log}\mathbf{h}= \mathbf{D}
\end{equation}
Where $ \bm{ \Omega}^{log} $ is called logarithmic spin introduced by Xiao and Bruhns\cite{xiao1997} with its explicit expression as:
\begin{equation}\label{eq:Log_spin}
\bm{\Omega}^{log} = \mathbf{W}+ \sum_{i \neq j}^{n}  \big(\frac{1+(\lambda_{i}/\lambda_{j})}{1-(\lambda_{i}/\lambda_{j})}+\frac{2}{\ln (\lambda_{i}/\lambda_{j})}\big) \mathbf{b}_i \mathbf{D} \mathbf{b}_j
\end{equation}
In which $\lambda_{i,j} (i,j=1,2,3) $ are the eigenvalues of Left Cauchy-Green tensor $ \mathbf{b} $ and $ \mathbf{b}_{i,j} $ are the corresponding subordinate eigenprojections. Given the antisymmetric logarithmic spin tensor, the associated second order rotation tensor $\mathbf{R}^{log}$ can be calculated through the following differential equation, and in most situations the initial condition can be assumed as $\mathbf{R}^{log}|_{t=0}= \mathbf I$.
\begin{equation}\label{eq:R_def}
{\bm{\Omega}^{log}}=\dot{\mathbf{R}}^{log}(\mathbf{R}^{log})^T
\end{equation} 
%
Following the definition of corotational integration by Khan and Huang\cite{khan1995continuum}, applying the logarithmic corotational integration procedure on Eqn.(\ref{eq:Log_strain_rate}), assuming initial conditions $\mathbf{h}|_{t=0}= \mathbf 0$, yields the total logarithmic strain as follows, 
\begin{equation}\label{eq:h-D}
\mathbf{h} = \int_{\text{coro.}}\mathbf{D} ~\text{d}t =(\mathbf{R}^{log})^T \bigg (\int \mathbf{R}^{log}\mathbf{D}^{e} (\mathbf{R}^{log})^T\text{d}t \bigg) \mathbf{R}^{log}  
\end{equation}
%

\subsection*{Additive Decomposition of Logarithmic Strain }\label{AdditiveStrain}
Starting from the additive decomposition of the total rate of deformation tensor $\mathbf D$ into an elastic part $\mathbf{D}^{e}$ plus a transformation part $\mathbf{D}^{tr}$,
\begin{equation}\label{eq:add_D}
\mathbf{D}=\mathbf{D}^{e}+\mathbf{D}^{tr}
\end{equation}
By virtue of Eqn.(\ref{eq:Log_strain_rate}) , elastic part $\mathbf{D}^{e}$ and transformation part  $\mathbf{D}^{tr}$  in Eqn.(\ref{eq:add_D})  can be rewritten as the $\mathring{\mathbf{h}}^{e\_log}$ and $\mathring{\mathbf{h}}^{tr\_log}$ respectively,
\begin{equation}\label{eq:add_h_rate1}
\mathring{\mathbf{h}}^{e\_log}=\mathbf{D}^{e};~~\mathring{\mathbf{h}}^{tr\_log}=\mathbf{D}^{tr}
\end{equation}
Combine Eqn.(\ref{eq:add_D}) and Eqn.(\ref{eq:add_h_rate1}) , the following equation can be obtained.
\begin{equation}\label{eq:add_h_rate2}
\mathring{\mathbf{h}}^{log}=\mathring{\mathbf{h}}^{e\_log}+\mathring{\mathbf{h}}^{tr\_log}
\end{equation}
Similar as Eqn.(\ref{eq:h-D}), applying logarithmic corotational integration procedure on Eqn.(\ref{eq:add_h_rate2}) yields,
\begin{subequations}
	\begin{align}
	\mathbf{h}^{e}  = \int_{\text{corot.}}\mathbf{D}^{e} ~\text{d}t =(\mathbf{R}^{log})^T \bigg(\int \mathbf{R}^{log}\mathbf{D}^{e} (\mathbf{R}^{log})^T\text{d}t \bigg ) \mathbf{R}^{log}  \\
	\mathbf{h}^{tr} = \int_{\text{corot.}}\mathbf{D}^{tr} ~\text{d}t =(\mathbf{R}^{log})^T \bigg (\int \mathbf{R}^{log}\mathbf{D}^{tr} (\mathbf{R}^{log})^T\text{d}t \bigg ) \mathbf{R}^{log}
	\end{align}
\end{subequations}
Thus the following additive decomposition of total logarithmic strain can be received. Namely, the total logarithmic strain can be additively split into an elastic part plus a transformation part. 
\begin{equation}\label{eq:add_h}
\mathbf{h}=\mathbf{h}^{e}+\mathbf{h}^{tr}
\end{equation}

\section*{MODEL FORMULATION}
\subsection*{Thermodynamic Potential of Constitutive Model}
In this section, the development for finite strain constitutive modeling of SMAs is going to be presented. This model formulation is based on the early established SMAs model by Lagoudas and coworkers \cite{boyd1996,lagoudas2012,lagoudas2008} within infinitesimal strain regime. We begin with an explicit expression for Gibbs free energy, in which Kirchhoff stress tensor $\bm{\uptau}$ and temperature $T$ are chosen as independent state variables for Gibbs free energy $G$. The martensitic volume fraction $ \xi $ and the second order transformation strain tensor $\mathbf{h}^{tr}$ are chosen as a set of internal state variables $ \bm{\Upsilon}=\{ \xi,\mathbf{h}^{tr}\} $ to capture the nonlinear response exhibited by polycrystalline SMAs. The explicit Gibbs free energy is given as:
\begin{equation}\label{eq:GIBBS_explicit}
\begin{aligned}
G=  -\dfrac{1}{2 \rho_{0}} \bm{\uptau} : \mathcal{S}\bm{\uptau} - \dfrac{1}{\rho_{0}}  \bm{\uptau} :[~\bm{\alpha}(T-T_0)+\mathbf{h}^{tr}]  +c \Big[(T-T_0)\\-T\ln (\dfrac{T}{T_0}) \Big] -s_0(T-T_0)+u_0+ \dfrac{1}{\rho_{0}}f(\xi)
\end{aligned}
\end{equation}
In which, $\mathcal{S}$ is the effective fourth-order compliance tensor calculated by using rule of mixtures defined in Eqn.(\ref{eq:S_mix}), $\mathcal{S}^A$ is compliance matrix for austenite phase, $\mathcal{S}^M$ is compliance matrix for martensite phase, and $\Delta\mathcal{S}$ is the phase difference between them. $ \bm{\alpha}$ is the second order thermoelastic expansion tensor, $ c $ is effective specific heat, $ s_0$ and $ u_0 $ are effective specific entropy and effective specific internal energy at reference state, respectively. Similar to the definition of effective compliance tensor $\mathcal{S}$ in Eqn.(\ref{eq:S_mix}), all those effective material parameters are using rule of mixture for their definition; $ T $ represents the temperature at current state while $ T_0 $ is temperature at reference state. $ f(\xi) $ is a smooth hardening function as it is introduced in original infinitesimal model.
\begin{subequations}
	\begin{align}\label{eq:S_mix}
	\mathcal{S}(\xi) &=\mathcal{S}^A + \xi(\mathcal{S}^M-\mathcal{S}^A)=\mathcal{S}^A + \xi\Delta\mathcal{S}  \\
	\bm{\alpha}(\xi) &=\bm{\alpha}^A + \xi(\bm{\alpha}^M-\bm{\alpha}^A)=\bm{\alpha}^A + \xi\Delta\bm{\alpha}  \\
	{c}(\xi)         &= c^A ~+ \xi(c^M-c^A)=c^A + \xi\Delta c  \\
	{s}_0(\xi)&= {s}_0^A + \xi({s}_0^M-{s}_0^A)={s}_0^A + \xi\Delta {s}_0\\\label{eq:u0}
	{u}_0(\xi)&= {u}_0^A + \xi({u}_0^M-{u}_0^A)={u}_0^A + \xi\Delta {u}_0
	\end{align}
\end{subequations}
%
Smooth hardening function $f(\xi)$ is defined in Eqn.(\ref{eq:Smooth_hardeing}) to consider the hardening effects associated with the transformation process. $a_1,a_2,a_3$ are introduced as intermediate material parameters in hardening function, and $n_1,n_2,n_3,n_4$ are curve fitting parameters to treat the smooth transition in the material response curve corner.
\begin{equation}\label{eq:Smooth_hardeing}
\begin{aligned}
f(\xi)=   \begin{cases} \cfrac{1}{2} a_1\Big(  \xi  + \frac{\xi^{n_1+1}} {n_1+1}+ \frac{(1-\xi)^{n_2+1}} {n_2+1} \Big)+a_3\xi ~, \; \dot{\xi}>0, \vspace{5pt} \\ 
\cfrac{1}{2} a_2\Big(  \xi  + \frac{\xi^{n_3+1}} {n_3+1}+ \frac{(1-\xi)^{n_4+1}} {n_4+1} \Big)-a_3\xi ~, \; \dot{\xi}<0 \end{cases}\\
\end{aligned} 
\end{equation}
Following standard Coleman-Noll procedure, the constitutive relation between stress and strain is derived as,  
\begin{equation}\label{eq:h_Cons_f}
\mathbf{h} = - \rho_{0}\frac{\partial G}{\partial \bm\uptau}=\mathcal{S}{\bm\uptau}+\bm\alpha(T-T_0)+ \mathbf{h}^{tr}
\end{equation}   
Constitutive relation between entropy $s$ and temperature $T$ can also be derived as,
\begin{equation}\label{eq:entropy_Cons_f}
s =- \rho_{0}\frac{\partial G}{\partial T}=\dfrac{1}{\rho_{0}} \bm{\uptau}:\bm\alpha+c\ln (\dfrac{T}{T_0}) + s_0
\end{equation}
The strict form dissipation inequality can be rewritten in terms of internal state variables $ \bm{\Upsilon}=\{ \xi,\mathbf{h}^{tr}\} $ as:
\begin{equation}\label{eq:Dissipation_strict_V2}
-\rho_{0}  \frac{\partial G}{\partial \mathbf{h}^{tr}} :\mathring{\mathbf{h}}^{tr} -\rho_{0}  \frac{\partial G}{\partial \xi}\dot{\xi} \geqslant 0
\end{equation}
\subsection*{Evolution Equation of Internal State Variables}\label{sec:Evolution}
Evolution equations between internal state variables is going to be set up in this part. Following the same assumption from the early work of Lagoudas and coworkers \cite{boyd1996,lagoudas2008,lagoudas2012}, the following evolution relationship between $ \mathbf{h}^{tr} $ and $\xi$ is proposed. It is worth to point out that the rate adopted here is not a conventional time rate but logarithmic rate instead.
\begin{equation}\label{eq:Trans_Evol}
{\mathring{\mathbf h}}^{\textit{tr}}= \bm{\Lambda}  \dot{\xi},  \ \  \bm\Lambda=\begin{cases}\bm{\Lambda}^{\textit{fwd}}, \; \dot{\xi}>0, \vspace{5pt} \\ \bm{\Lambda}^{\textit{rev}}, \; ~\dot{\xi}<0, \end{cases}\\
\end{equation}
Where $\bm\Lambda^{\textit{fwd}}$ is called forward transformation direction tensor while $\bm\Lambda^{\textit{rec}}$ is called reverse transformation direction tensor. They are defined as follows respectively:
\begin{equation}\label{eq:direction}
\bm\Lambda^{\textit{fwd}}=
\frac{3}{2} H^{\textit{cur}} 
\frac{\bm{\uptau}^{'}}{\bar{\uptau}^{'}},  \ \bm\Lambda^{rev}=
\frac{\mathbf h^{\textit{tr-r}}}{{\xi}^{\textit{r}}}.
\end{equation}
In Eqn.(\ref{eq:direction}), $ \bm{\uptau}^{'} $ is the deviatoric part of Kirchhoff stress tensor which is calculated by {\small $ \bm{\uptau}^{'} =\bm{\uptau} -{\small \frac{1}{3}}\textrm{tr}(\bm{\uptau})~\mathbf{I} $}, where $ \mathbf{I} $ is the second order identity tensor. The effective (von Mises equivalent) stress is given by $ \bar{\uptau}^{'} ={\small \sqrt{{{\small \frac{3}{2}}\bm\uptau}^{'}:\bm{\uptau}^{'}}}$. And $\mathbf h^{\textit{tr-r}}$ and ${\xi}^{r}$ represents the transformation strain value and martensitic volume fraction value at the starting point of reverse transformation. From experimental observation, the transformation strain is not a constant value but it usually depends on the material current stress state, thereby a current transformation strain $H^{\textit{cur}}$ is introduced through a exponential function dependent on current stress state in Eqn.(\ref{eq:Hcur}), where $H^{\textit{max}}$ is the maximum transformation strain and $\textit{k}_t$ is a curve fitting parameter, ${ \bar{\bm\uptau}}$ is effective stress of Kirchhoff stress.
%
\begin{equation}\label{eq:Hcur}
H^{\textit{cur}}(\bm\uptau)= H^{\textit{max}}(1-e^{-\textit{k}_t { \bar{\bm\uptau}}})
\end{equation}
\subsection*{Transformation Function}\label{Trans_Func}
After the definition of evolution equation for internal state variable, the next objective is to define a proper criterion to determine whether the transformation will happen or not. Substitute the evolution Eqn.(\ref{eq:Trans_Evol}) into the strict form dissipation inequality Eqn.(\ref{eq:Dissipation_strict_V2}), we obtain the following equation:
\begin{equation}\label{eq:Dissipation_xi}
(\bm\uptau:\bm\Lambda-\rho_{0}  \frac{\partial G}{\partial \xi})\dot{\xi}=\pi\dot{\xi}\geqslant 0 
\end{equation}
Scalar variable $\pi$ is called general thermodynamic driving force conjugated to martensitic volume fraction $\xi$. Substitution of Gibbs free energy $G$ in Eqn.(\ref{eq:GIBBS_explicit}) into Eqn.(\ref{eq:Dissipation_xi}) yields the expression for $ \pi $:
\begin{equation}\label{eq:Driving_Force}
\begin{aligned}
\pi(\bm\uptau,T,\xi)=\bm\uptau:\bm\Lambda+
\dfrac{1}{2}\bm\uptau:{\Delta}\mathbf{S}:\bm\uptau+\bm\uptau:{\Delta}\mathbf{\alpha}(T-T_0)-\rho_0\Delta c  
\big[ T-T_0\\ -T\ln(\dfrac{T}{T_0}) \big ] + \rho_0\Delta s_0 T - \rho_0\Delta u_0 - \frac{\partial f}{\partial \xi}
\end{aligned}
\end{equation}
Where material parameters $\Delta \mathcal{S},\Delta \mathbf{\alpha}, \Delta c, \Delta s_0, \Delta u_0 $ have the definition from Eqn.(\ref{eq:S_mix}) to Eqn.(\ref{eq:u0}). We assume that whenever the thermodynamic driving force $\pi$ reaches a critical value $Y $ ($ -Y $), the forward (reverse) phase transformation will take place. Therefore, a transformation function $\Phi$, defined as Eqn.(\ref{eq:Transfor_Fun}), can be used as a criteria to determine the occurrence for phase transformation. 
\begin{equation}\label{eq:Transfor_Fun}
\normalfont{\Phi}=\begin{cases}~~\pi - Y, \; \dot{\xi}>0, \vspace{5pt} \\ -\pi - Y, \; \dot{\xi}<0, \end{cases}\\
\end{equation}
In the continuous development of Lagoudas et al.\cite{lagoudas2012} model, a stress dependent critical value $Y$ was proposed. A constant reference value $Y_0$ and an additional parameter D are introduced into $Y$ such that it can better capture the smooth transition during transformation initial zone. Critical value $Y$ is defined as follows, 
\begin{equation}\label{Critical_Y}
Y(\bm{\uptau}) = \begin{cases}Y_0 + D\bm\uptau:\bm\Lambda^{\textit{fwd}}, \; \dot{\xi}>0, \vspace{5pt} \\ Y_0 + D\bm\uptau:\bm\Lambda^{\textit{rev}}, \; \dot{\xi}<0, \end{cases}\\
\end{equation}
To be satisfied with the principle of maximum dissipation, a so-called Kuhn-Tucker constraint conditions has also been placed on the evolution equation for internal state variables, which are expressed as follows for forward and reverse transformation cases respectively:
\begin{equation}\label{eq:Kuhn-Tucker}
\begin{aligned}
\dot{\xi} \geqslant 0; \quad \Phi(\bm\uptau,T,\xi)= ~~\pi - Y \leqslant 0;  \quad  \Phi\dot{\xi}=0;~~~\bf{(A\Rightarrow M)}\\
\dot{\xi} \leqslant 0; \quad \Phi(\bm\uptau,T,\xi)= -\pi - Y \leqslant 0; \quad   \Phi\dot{\xi}=0;~~~ \bf{(M\Rightarrow A)}
\end{aligned}
\end{equation}
%
%
%
\section*{Consistent Tangent Stiffness and Thermal Matrix}\label{sec:Jacobian}
In this section, a detailed derivation of consistent tangent stiffness matrix and thermal matrix is provided to complete the proposed model. In most typical displacement-based commercial finite element software, like Abaqus, when an user defined material subroutine (UMAT) is written, consistent tangent matrices are usually required from finite element program solver to achieve a fast and accurate solution for global equilibrium equations during Newton-Raphson iteration procedure. Normally, consistent tangent matrices can be expressed in a rate form as defined in Eqn.(\ref{eq:Jacobian}), $\mathcal{L}$ is consistent tangent stiffness matrix and $\Theta$ is consistent thermal matrix.
\begin{equation}\label{eq:Jacobian}
\mathring{\bm{\uptau}}  =  {\mathcal{L}}\mathring{\mathbf h} 
+ \Theta \dot T
\end{equation}
Apply logarithmic rate on constitutive stress strain Eqn.(\ref{eq:h_Cons_f}), where $\mathcal{C}$ is the stiffness matrix, the inverse of compliance tensor $\mathcal{S}$ 
\begin{equation}\label{eq:rate_cons}
\mathring{\bm{\uptau}}  = {\mathcal{C}}~[\mathring {\mathbf h}  -  \bm\alpha\dot T - ({\Delta\mathcal{S}}\bm{\uptau} +  \Delta{\bm\alpha}(T-T_0) + \bm\Lambda )\dot \xi~]
\end{equation}
Take chain rule differentiation on transformation function Eqn.(\ref{eq:Transfor_Fun}), and replace conventional time rate with logarithmic rate. Be noticed that logarithmic rate is equivalent to conventional rate when it is applied on a scalar.
\begin{equation}\label{eq:rate_TransFun}
\dot{\Phi}  = \partial_{\bm{\uptau}}\Phi:\mathring{\bm\uptau} + \partial_{T}\Phi\dot{T} + \partial_{\xi}\Phi\dot{\xi} = 0  
\end{equation}
Substitute Eqn.(\ref{eq:rate_cons}) back into Eqn.(\ref{eq:rate_TransFun}) to eliminate $\mathring{\bm{\uptau}}$ and solve it for $\dot \xi$, the following expression for $\dot \xi$ can be obtained,
\begin{equation}\label{eq:rate_xi}
\dot{\xi}  = -\dfrac{\partial_{\bm\uptau}\Phi: \mathcal{C} \mathring {\mathbf h} +(\partial_{T}\Phi- \partial_{\bm\uptau}\Phi: \mathcal{C} \bm \alpha)\dot T}{\partial_{\xi}\Phi- \partial_{\bm\uptau}\Phi: \mathcal{C} (\Delta\bm S\bm\uptau+\bm\Lambda+\bm)} 
\end{equation}
Substitute Eqn.(\ref{eq:rate_xi}) back into rate form constitutive Eqn.(\ref{eq:rate_cons}), eliminate $\dot \xi$ and after some tensorial manipulations, a final explicit expression corresponding to Eqn.(\ref{eq:Jacobian}) can be expressed as follows,
\begin{equation}\label{eq:}
\begin{aligned}
\mathring{\bm{\uptau}}  = \Big[ \mathcal{C}+\dfrac{[\mathcal{C}({\Delta\mathcal{S}}\bm{\uptau}+ \bm\Lambda )] \otimes  [\mathcal{C} \partial_{\bm\uptau}\Phi]}{\partial_{\xi}\Phi- \partial_{\bm\uptau}\Phi: \mathcal{C}(\Delta\mathcal{S}\bm\uptau+\bm\Lambda)} \Big] \dot{\bm\varepsilon} 
+  \Big[ 
- \mathcal{C}\bm\alpha \\+
\dfrac{ \mathcal{C}({\Delta\mathcal{S}}\bm{\uptau}+ \bm\Lambda)  (\partial_{T}\Phi - \partial_{\bm\uptau}\Phi: \mathcal{C}\bm\alpha )}
{\partial_{\xi}\Phi- \partial_{\bm\uptau}\Phi: \mathcal{C}(\Delta\mathcal{S}\bm\uptau+\bm\Lambda)} 
\Big] \dot T 
\end{aligned}
\end{equation}
In which consistent tangent stiffness matrix ${\mathcal{L}}$ is obtained as, 
\begin{equation}
{\mathcal{L}}=\mathcal{C}+\dfrac{[\mathcal{C}({\Delta\mathcal{S}}\bm{\uptau}+ \bm\Lambda ] \otimes  [\mathcal{C} \partial_{\bm\uptau}\Phi]}{\partial_{\xi}\Phi- \partial_{\bm\uptau}\Phi: \mathcal{C}(\Delta\mathcal{S}\bm\uptau+\bm\Lambda)} 
\end{equation}
and consistent thermal matrix $\Theta$ is derived as, 
\begin{equation}
\Theta = 
- \mathcal{C}\bm\alpha + 
\dfrac{ \mathcal{C}({\Delta\mathcal{S}}\bm{\uptau}+ \bm\Lambda )  (\partial_{T}\Phi - \partial_{\bm\uptau}\Phi: \mathcal{C}\bm\alpha )}
{\partial_{\xi}\Phi- \partial_{\bm\uptau}\Phi: \mathcal{C}(\Delta\mathcal{S}\bm\uptau+\bm\Lambda)} 
\end{equation}


\section*{NUMERICAL RESULTS}\label{Result}
In this section, Boundary value problems, including a simple extension problem for SMA bar and a shear problem for a SMA torque tube,  will be investigated to test the capabilities of proposed finite strain model, the response predicated by which is going to be compared with its infinitesimal counterpart. Besides, the cyclic pseudoelastic response of the SMA torque tube is also going to be examined to evaluate the effects of artificial residual stress introduced by objective rates. 

\subsection*{Bar Problem}\label{subsec:bar}
\begin{figure*}[t!]
	\centering
	\includegraphics[width=0.6\textwidth]{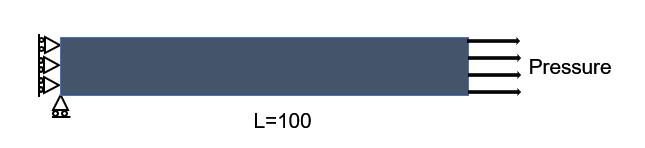}
	\caption{\uppercase{Schematic of a SMA bar extension problem under external pressure at constant temperature}}
	\label{fig:Bar_Schematic}
	\vspace{0.5cm}
\end{figure*}

The first boundary value problem analyzed is a SMA bar experiencing stress-induced phase transformation under external pressure at constant temperature, which is typically called a pseudoelastic (or superelastic) response exhibited by SMAs. Referring to Fig.\ref{fig:Bar_Schematic} for problem schematic, a long SMA bar with length  $L=100$ unit (mm) and has a square cross section with edge length $ a=10 $ unit (mm). Mechanical boundary conditions are left face fixed in certain lines and points to remove rigid body motions and right face is free to move but with an applied external pressure in its length direction. The material parameters used in this simulation are summarized in Tab.\ref{tab:MaterialProperty_bar} referenced from literature \cite{lagoudas2008}.

The loading history of SMA bar is the following: increase the pressure on right face from zero value to a maximum value 1200 (MPa), during which the SMA bar will experience a stress-induced extension due to forward transformation from austenite phase to detwinned martensite phase; Decrease the pressure linearly from its maximum value to zero, during which the SMA bar will contract to its original length due to the reverse transformation from detwinned martensite phase to austenite phase. The temperature is kept constant as $360$K throughout the whole loading procedure. The material response predicted by proposed finite strain model are compared against its infinitesimal counterpart.

\begin{table*}[h!] 
	\centering
	\caption{SUMMARY OF USED MATERIAL PARAMETERS FOR NUMERICAL  SIMULATIONS}
	\renewcommand{\arraystretch}{0.8}
	\begin{tabular}{c|lr|ll} \toprule
		Type                         &Parameter                        & Value                                   &Parameter            & Value  \\                                       \midrule
		&$E_A$                            & 90   [GPa]                              & $C_A$                & 16  [MPa/K]\\
		&$E_M$                            & 63   [GPa]                              & $C_M$               & 10  [MPa/K]\\
		Material Constants       &$\nu_A=\nu_M$                          & 0.3~~~~~~~~                              & $M_s$                & 308  [K]\\
		10                      &$\alpha_A=\alpha_M$    &    2.2$\times$10$^{-5}$ [K$^{-1}$]                           & $M_f$                &242  [K]\\
		&                       &           & $A_s$                 &288  [K]\\
		&                       &       & $A_f $                 & 342  [K]\\                                   \midrule
		
		&$ H^\textit{max}$               & 1\%, 3\%, 5\%, 10\%                     & $n_1$              & 0.5 \\
		Smooth Hardening         & $k_t$                          &0.0075                              & $n_2$         & 0.5 \\
		6                               &              &                     & $n_3$         &  0.5\\                             
		& & & $n_4$     &  0.5           \\  
%
		\bottomrule
	\end{tabular}
	\label{tab:MaterialProperty_bar}
\end{table*}
\begin{figure}[t]
	\centering\hspace*{-1cm}
	\includegraphics[width=0.5\textwidth]{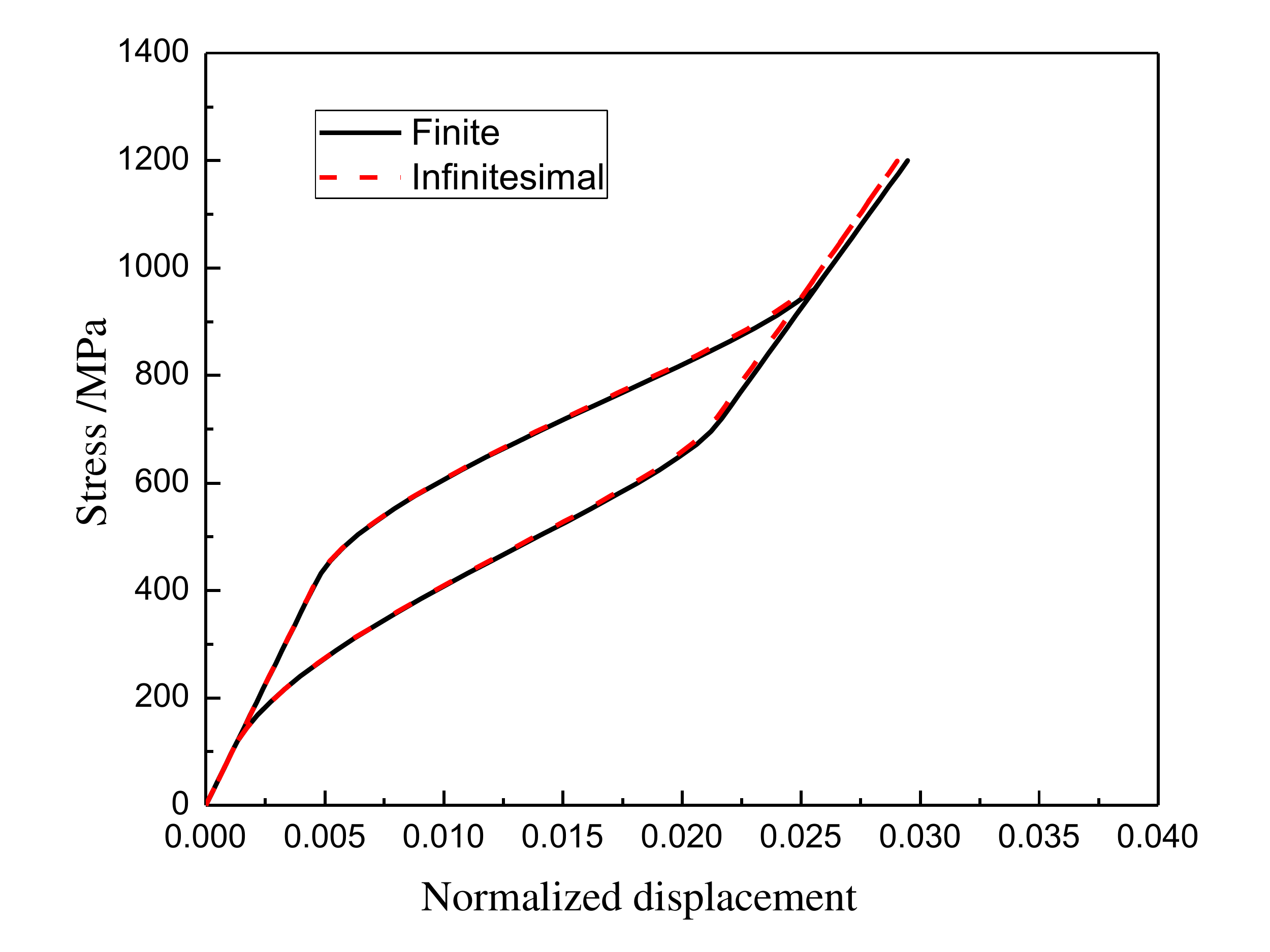}\vspace{-0.3cm}
	\caption{\uppercase{Comparison of pseudoelastic response for a SMA bar predicted by  proposed finite strain model and infinitesimal strain model with transformation strain} ${H}^{max}=1\%$ }
	\label{fig:Bar_H01}
\end{figure}
\begin{figure}[t]
	\centering\hspace*{-1cm}
	\includegraphics[width=0.5\textwidth]{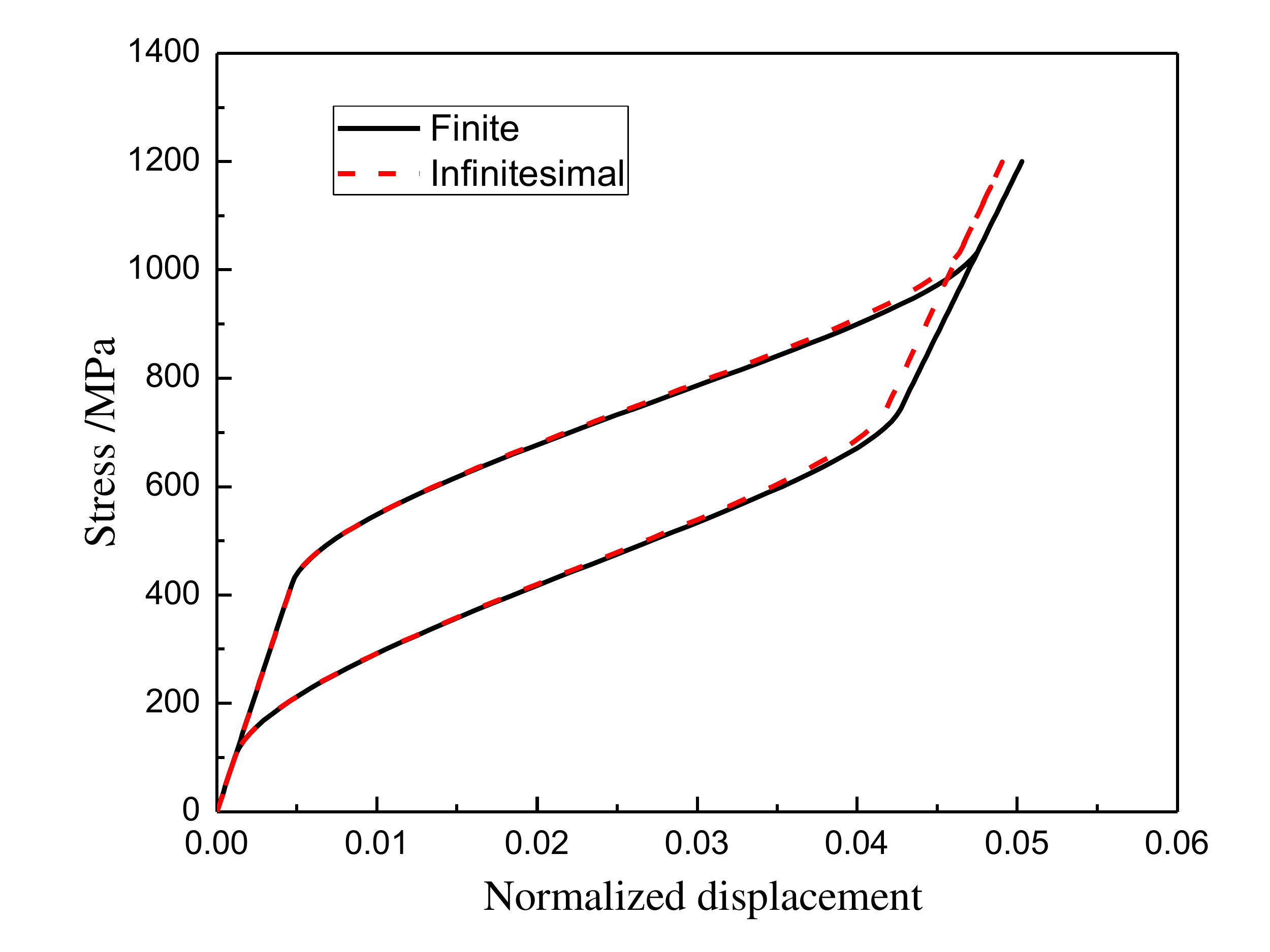}\vspace{-0.3cm}
	\caption{\uppercase {Comparison of pseudoelastic response for a SMA bar predicted by  proposed finite strain model and infinitesimal strain model with transformation strain} ${H}^{max}=3\%$ }
	\label{fig:Bar_H03}
\end{figure}
\begin{figure}[t]
	\centering\hspace*{-1cm}
	\includegraphics[width=0.5\textwidth]{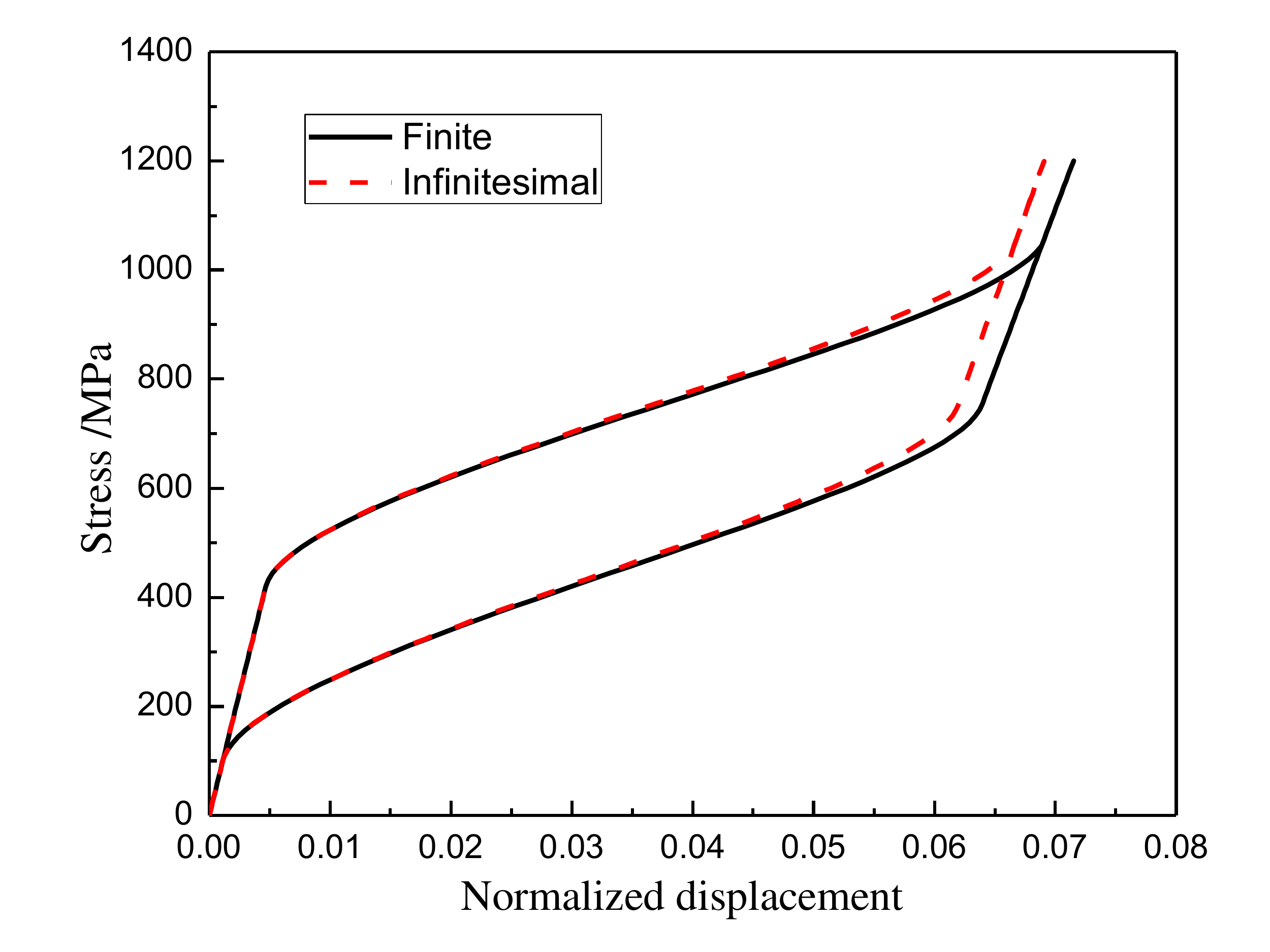}\vspace{-0.3cm}
	\caption{\uppercase {Comparison of pseudoelastic response for a SMA bar predicted by  proposed finite strain model and infinitesimal strain model with transformation strain} ${H}^{max}=5\%$ }
	\label{fig:Bar_H05}
\end{figure}
\begin{figure}[t]
	\centering\hspace*{-1cm}
	\includegraphics[width=0.5\textwidth]{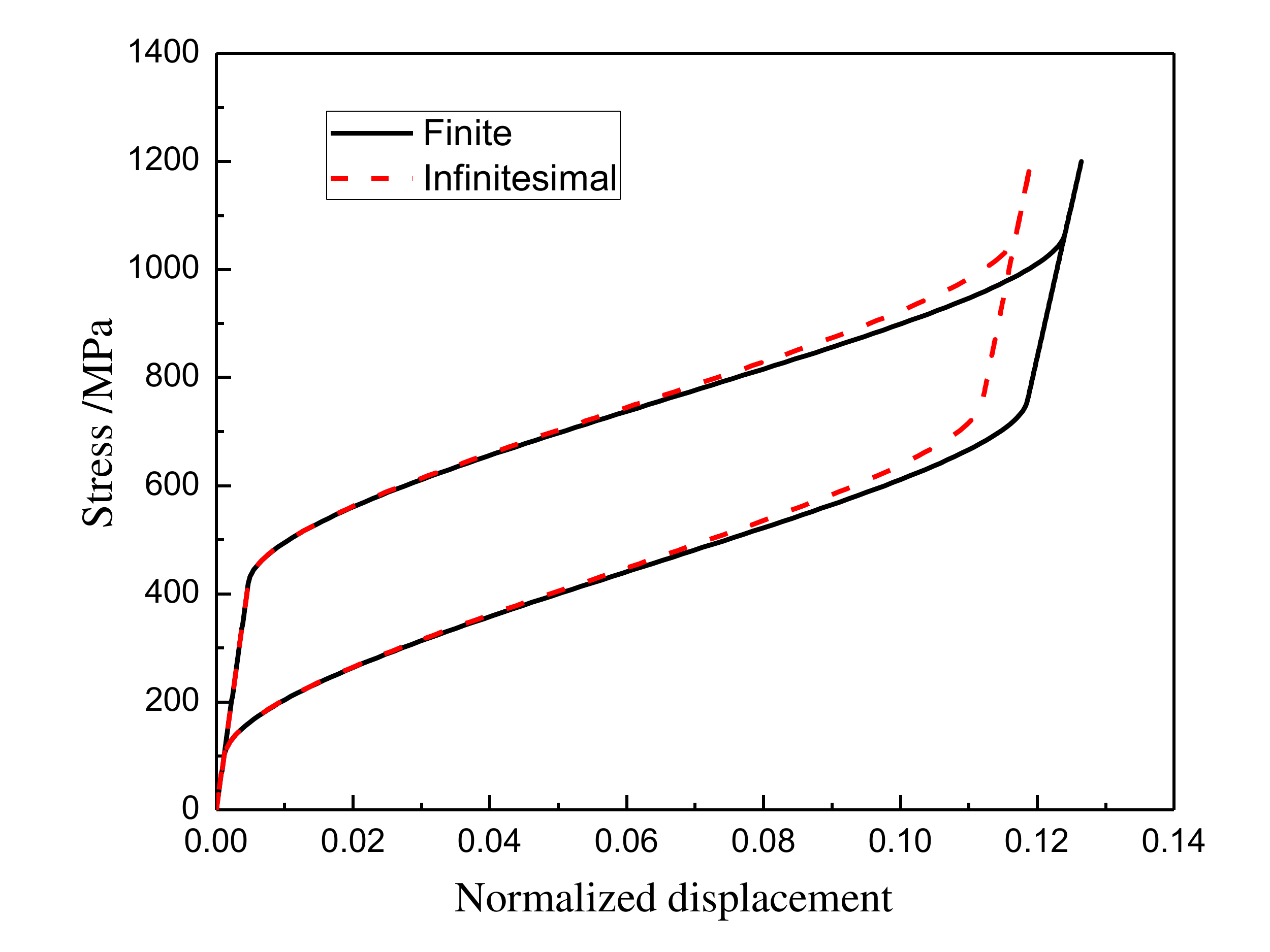}\vspace{-0.3cm}
	\caption{\uppercase {Comparison of pseudoelastic response for a SMA bar predicted by  proposed finite strain model and infinitesimal strain model with transformation strain} ${H}^{max}=10\%$ }
	\label{fig:Bar_H10}
\end{figure}
%
As the result is shown in Fig..\ref{fig:Bar_H01}, when the maximum transformation is strain $1\%$, the difference of pseudoelastic response curve predicted by the two models is almost negligible. However, with the increasing of maximum transformation strain from $1\%$ to $3\%$, and more to $5\%$ and $10\%$, the difference of predicted pseudoelastic response is becoming predominant. Specifically, with max transformation strain $H^{max}=3\%$ in Fig.\ref{fig:Bar_H03}, the difference is becoming perceivable at the end of forward transformation, normalized displacement has larger value predicted by proposed finite strain model in contrast to infinitesimal strain model. This response difference is growing substantial in the cases of $H^{max}$ to be $5\%$ and $10\%$. Since there is no rotation involved here, the difference for the result comes from the fact that infinitesimal strain neglect the higher order terms in strain measure while logarithmic strain used in proposed model doesn't.
%
\subsection*{Tube Problem}\label{sec:torque_tube}
\begin{figure*}[t]
	\centering
	\includegraphics[width=0.8\textwidth]{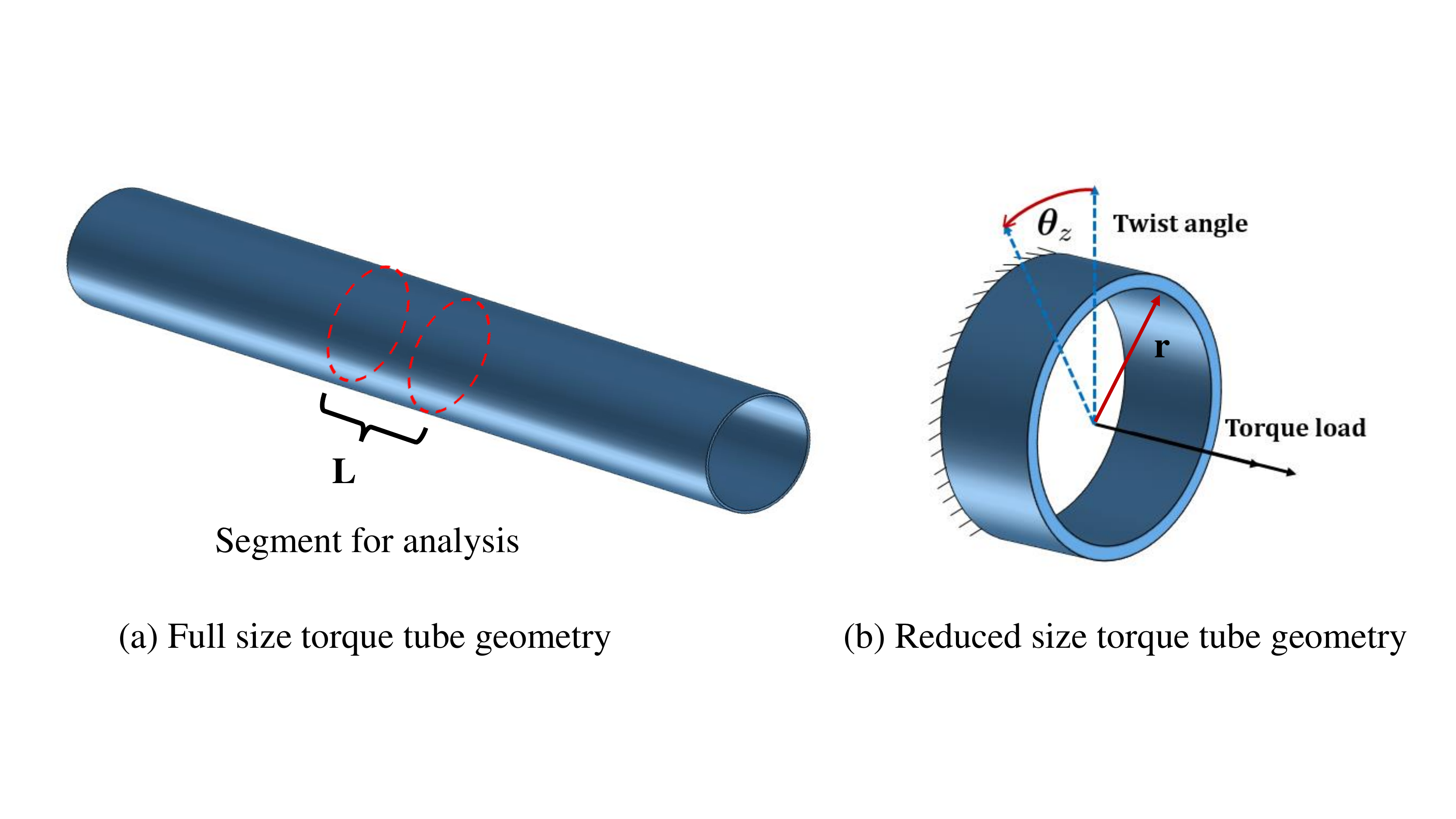}\vspace{-1.8cm}
	\caption{\uppercase{Schematic for the shear problem of hollow cylindrical SMA torque tube} }{\vspace{-0.2cm}}
	\label{fig:Tube_Schematic}
\end{figure*}

In this section, a shear problem of a hollow cylindrical SMA torque tube under torsion loading at constant temperature is studied. Referring to Fig.\ref{fig:Tube_Schematic} for problem schematic, a full size geometry of long hollow cylindrical torque tube is depicted in Fig.\ref{fig:Tube_Schematic}(a). In order to reduce the computational cost, only one small segment tube with length $L$ and inner radius $r$ is chosen for analysis. The segment tube has a geometry ratio of $L/r =2$. 

As it is shown in Fig.\ref{fig:Tube_Schematic}(b), the mechanical boundary condition is following: the tube left face is fixed in all degrees of freedom while the right face is subject to a torsion loading through a twist angle $\theta_z(t)$ along its longitudinal direction. The loading history is: the twist angle $\theta_z(t)$ increases proportionally from zero to a maximum value at $t = 0.5$, during which the torque tube undergoes a fully forward phase transformation from austenite phase to detwinned martensite phase. Afterwards, twist angle $\theta_z(t)$ proportionally decreases from its peak value to zero, during which the SMA tube experienced a reverse phase transformation from martenite phase to austenite phase. Temperature is kept as a constant value of $360$ K throughout this whole loading process. Material parameters used in this numerical experiment are choosing from Tab.\ref{tab:MaterialProperty_bar}.

\begin{figure}[t]
	\vspace{-0.5cm}\centering\hspace*{-0.8cm}
	\includegraphics[width=0.5\textwidth]{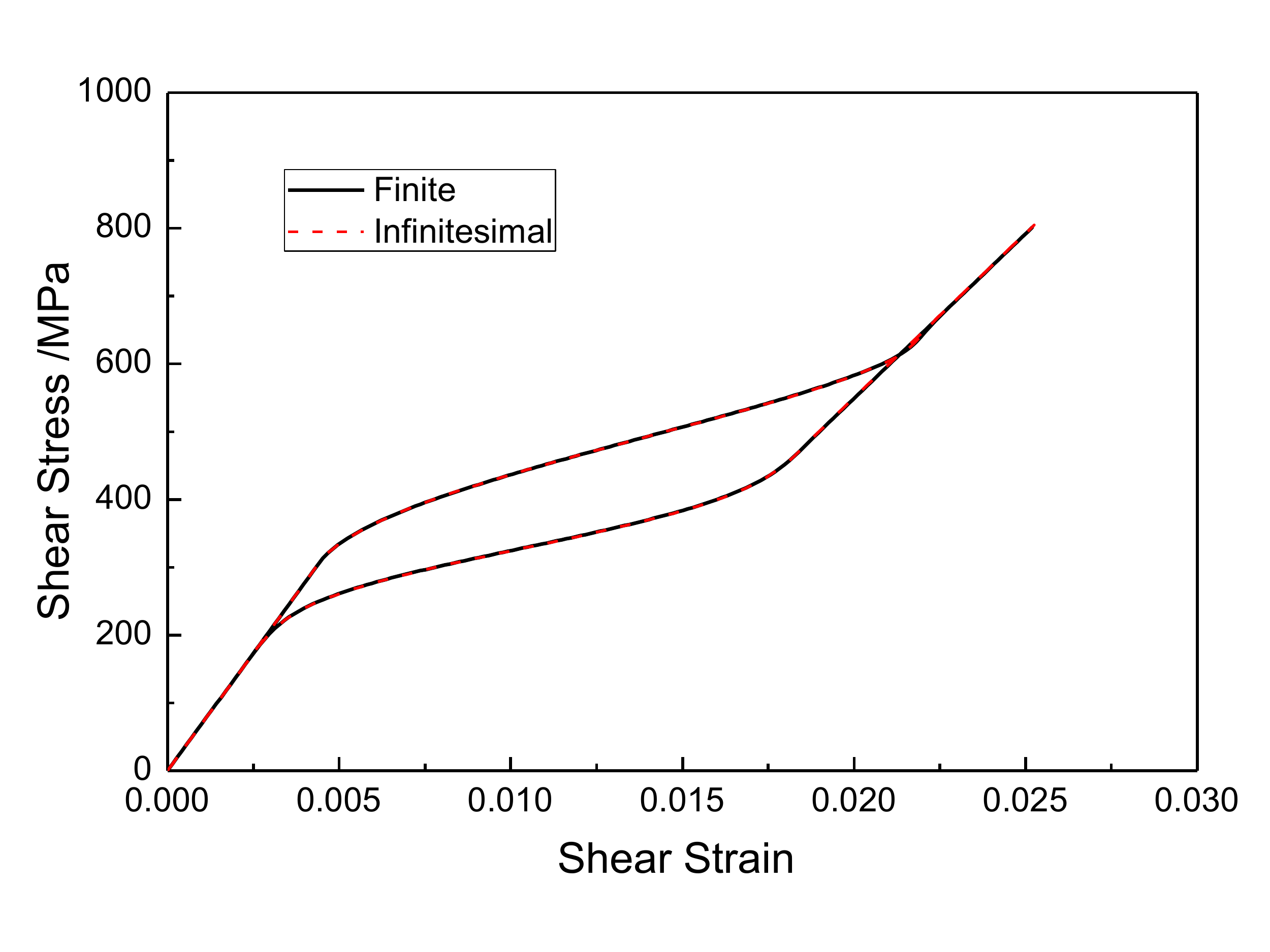}\vspace{-0.5cm}
	\caption{\uppercase{Comparison of pseudoelastic response for a short SMA torsion tube predicted by proposed finite strain and infinitesimal strain model with transformation strain} ${H}^{max}=1\%$, \uppercase{geometry ratio} $L/r=2$  }
	\label{fig:Tube_L6_H01}
\end{figure}
\begin{figure}[t]
	\vspace{-0.5cm}\centering\hspace*{-0.8cm}
	\includegraphics[width=0.5\textwidth]{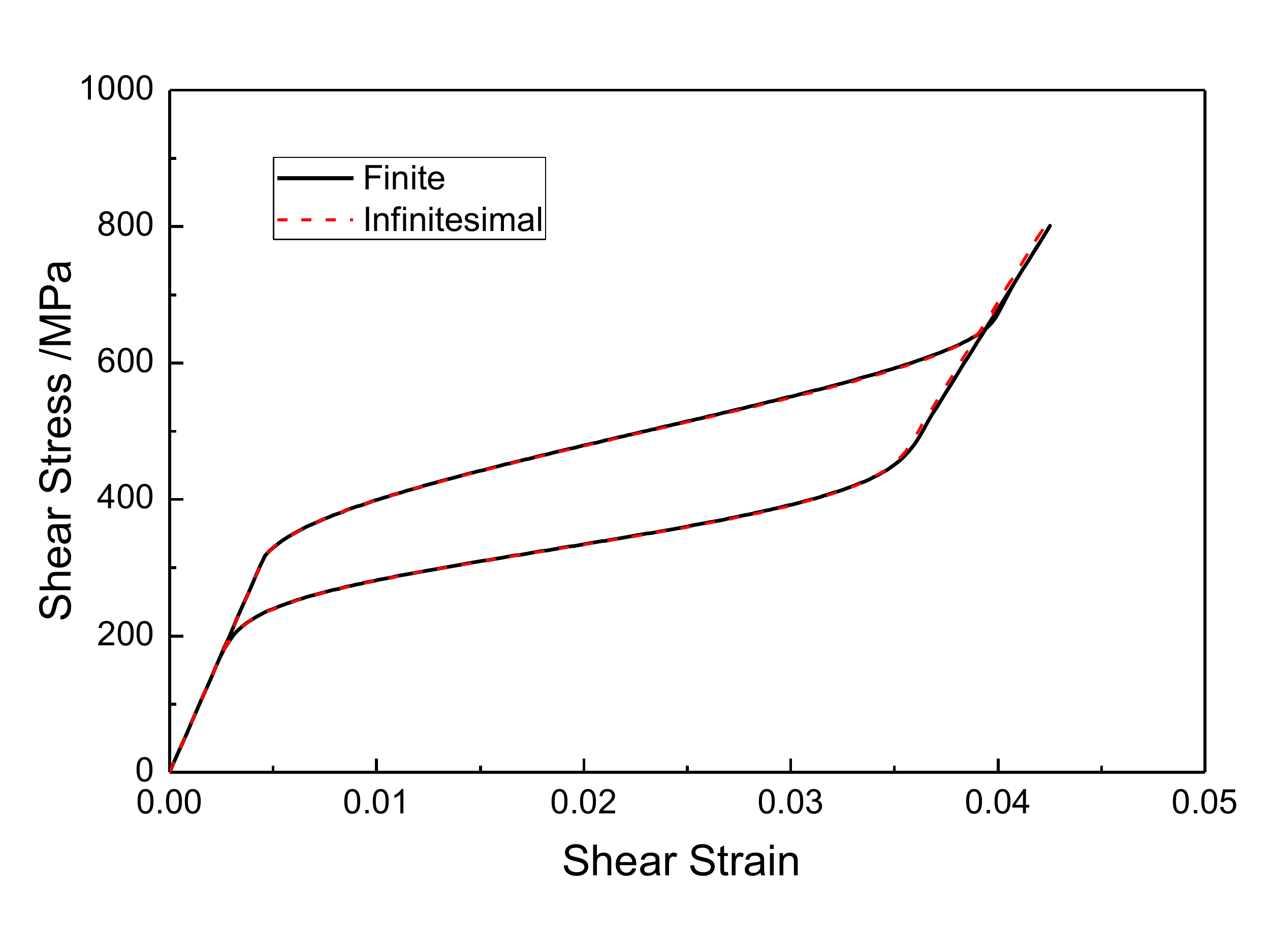}\vspace{-0.5cm}
	\caption{\uppercase{Comparison of pseudoelastic response for a short SMA torsion tube predicted by proposed finite strain and infinitesimal strain model with transformation strain} ${H}^{max}=3\%$, \uppercase{geometry ratio} $L/r=2$  }
	\label{fig:Tube_L6_H03}
\end{figure}
\begin{figure}[t]
	\vspace{-0.5cm}\centering\hspace*{-0.8cm}
	\includegraphics[width=0.5\textwidth]{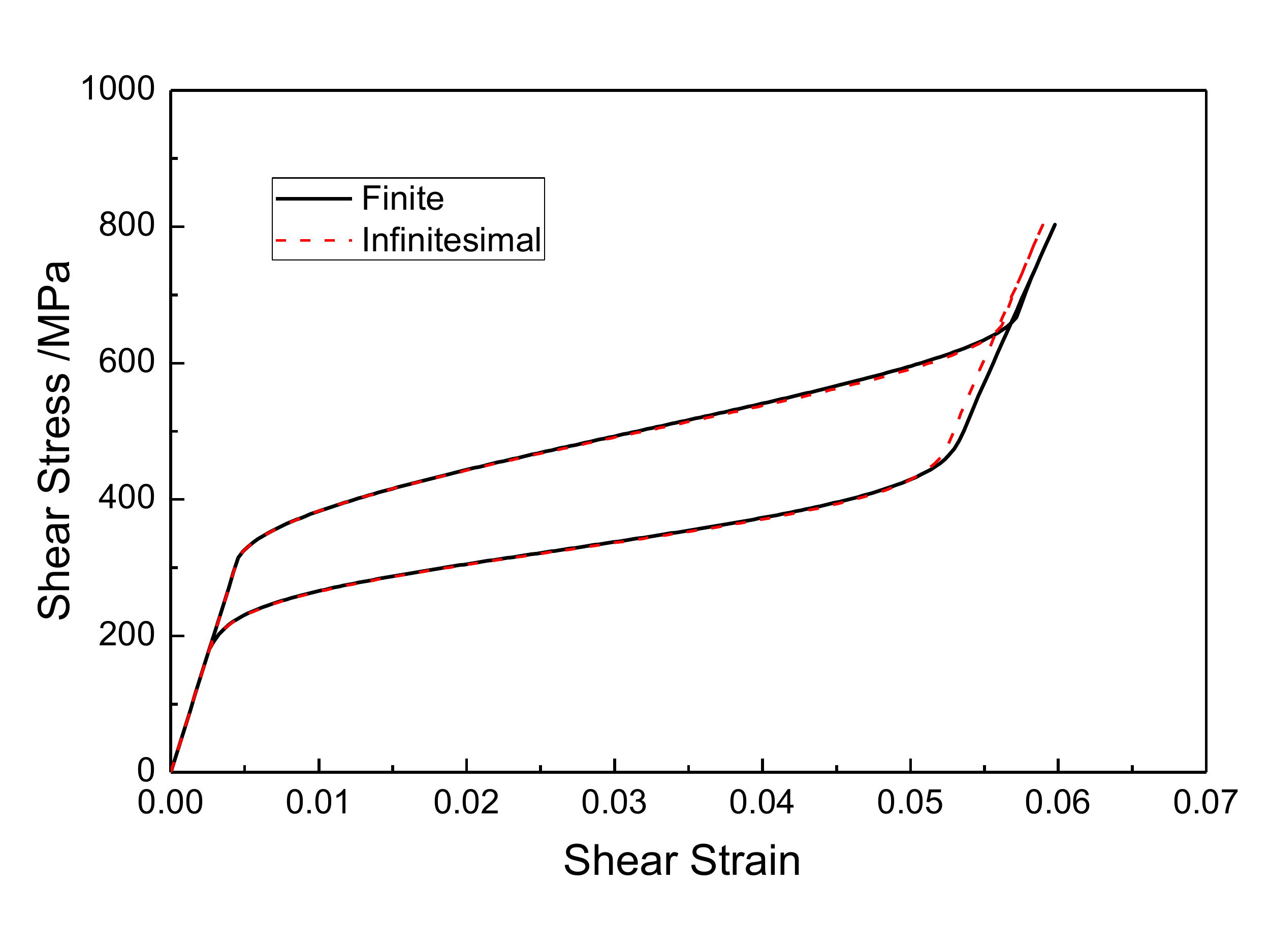}\vspace{-0.5cm}
	\caption{\uppercase{Comparison of pseudoelastic response for a short SMA torsion tube predicted by proposed finite strain and infinitesimal strain model with transformation strain} ${H}^{max}=5\%$, \uppercase{geometry ratio} $L/r=2$  }
	\label{fig:Tube_L6_H05}
\end{figure}
\begin{figure}[t]
	\vspace{-0.5cm}\centering\hspace*{-0.8cm}
	\includegraphics[width=0.5\textwidth]{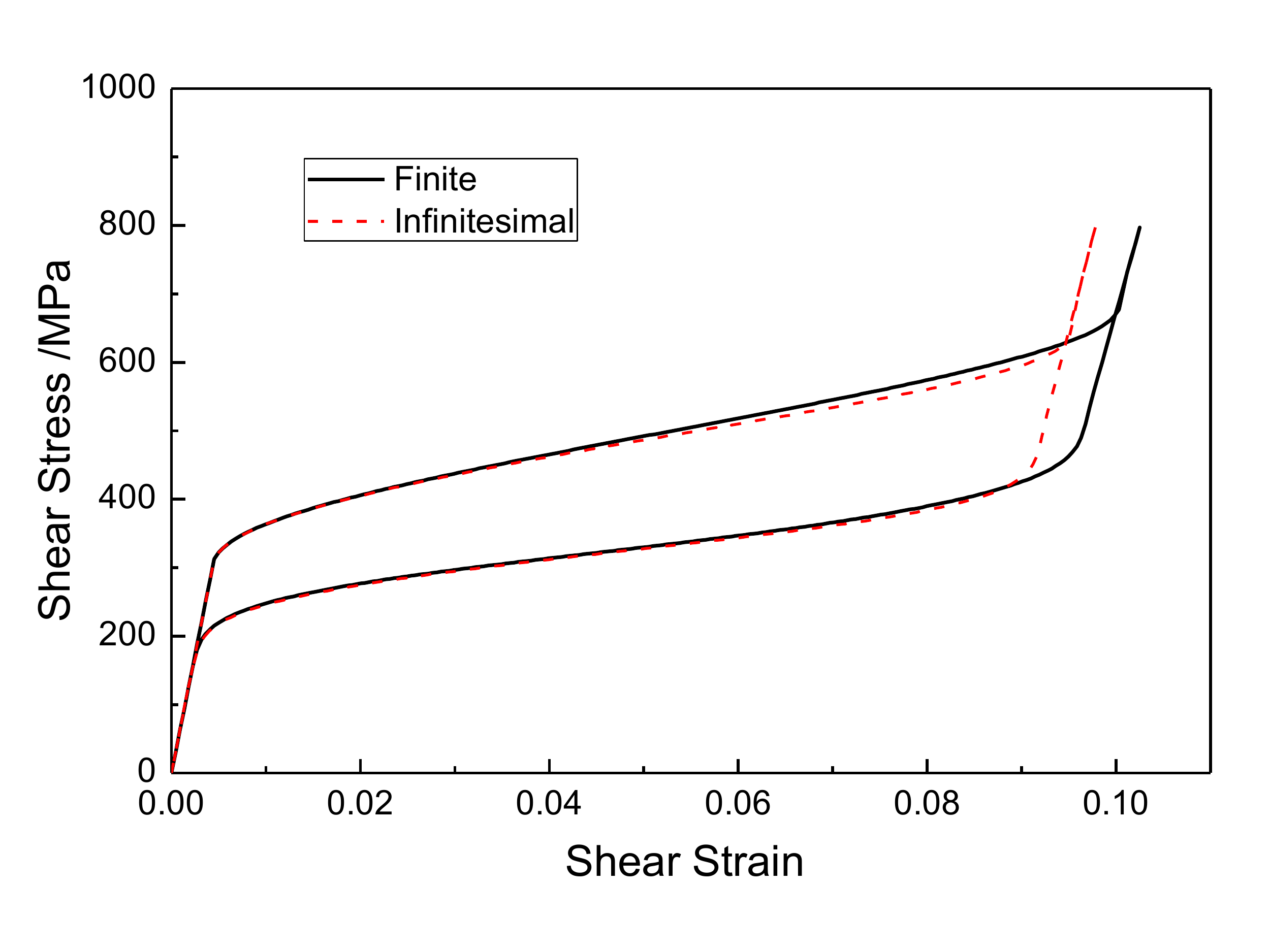}\vspace{-0.5cm}
	\caption{\uppercase{Comparison of pseudoelastic response for a short SMA torsion tube predicted by proposed finite strain and infinitesimal strain model with transformation strain} ${H}^{max}=10\%$, \uppercase{geometry ratio} $L/r=2$  }
	\label{fig:Tube_L6_H10}\vspace{-0.5cm}
\end{figure}
%
%
%
%
It can be observed from the results showing from Fig.\ref{fig:Tube_L6_H01} to Fig.\ref{fig:Tube_L6_H10}, although there is no considerable difference on the material response predicted by both models for maximum transformation strain $H^{max}= 1\% \sim 3\% $, perceivable difference for maximum transformation strain $H^{max}= 5\% $ is showing in Fig.\ref{fig:Tube_L6_H05}, and this difference continues to grow substantially as maximum transformation strain increases to $H^{max}= 10\% $ in Fig.\ref{fig:Tube_L6_H10}. Through the above comparison on the pseudoelastic response for segment torque tube, it is demonstrated that, under large rotations, the difference of predicted material response by proposed finite strain model and infinitesimal model will be perceivable within moderate strain regime of $5\%$, and it continues to grow substantially in large strain regime at $10\%$.


\section*{Cyclic Response of SMA Torque Tube }
As it was discussed in the section of introduction, based on the additive decomposition in finite deformation theory, a rate form hypoelastic constitutive equation is often seen as stress and strain relationship, in which a objective rate is  usually required on the stress tensor in order to meet the principle of objectivity. There are many different objective rates proposed by different researchers along this line, among which several famous ones are Zaremba-Jaumann rate, Green-Naghdi rate and Truesdell rate etc.. However, for a long period of time, rate form hypoelastic constitutive theory has been criticized for its inconsistent choices on objective rates\cite{simo2006}, and also the rate form hypoelastic constitutive equation is not able to be integrated to deliver a algebraic constitutive equation, thereby many spurious phenomenons, such as shear stress oscillation, dissipative phenomenon or artificial residual stress are observed in even simple elastic deformation.
The aforementioned self-inconsistent issues of hypoelastic constitutive model are resolved by the logarithmic rate proposed by Xiao et al. \cite{xiao1997,xiao1997hypo,xiao2006}, Bruhns et al.\cite{bruhns1999self,bruhns2001large,bruhns2001self}, Meyers et al.\cite{meyers2003elastic,meyers2006choice}. 

In this section, a segment torsion tube is analyzed by proposed finite strain model and its original infinitesimal counterpart with Abaqus NLGEOM (nonlinear geometry) option to extend it for large deformation analysis. The loading conditions and material parameters used are the same as previous segment torsion tube. It should be clear, when the Abaqus NLGEOM option is activated for infinitesimal strain based SMA model during implicit analysis, the strain measure is logarithmic strain and Jaumman rate is the utilized objective rate to capture rotation deformation. As it is mentioned about with objective rates, artificial residual stresses will be introduced by using other objective rates except for the logarithmic one. 

\begin{figure*}[t!]
	\centering\vspace{-0.2cm}
	\includegraphics[width=0.8\textwidth]{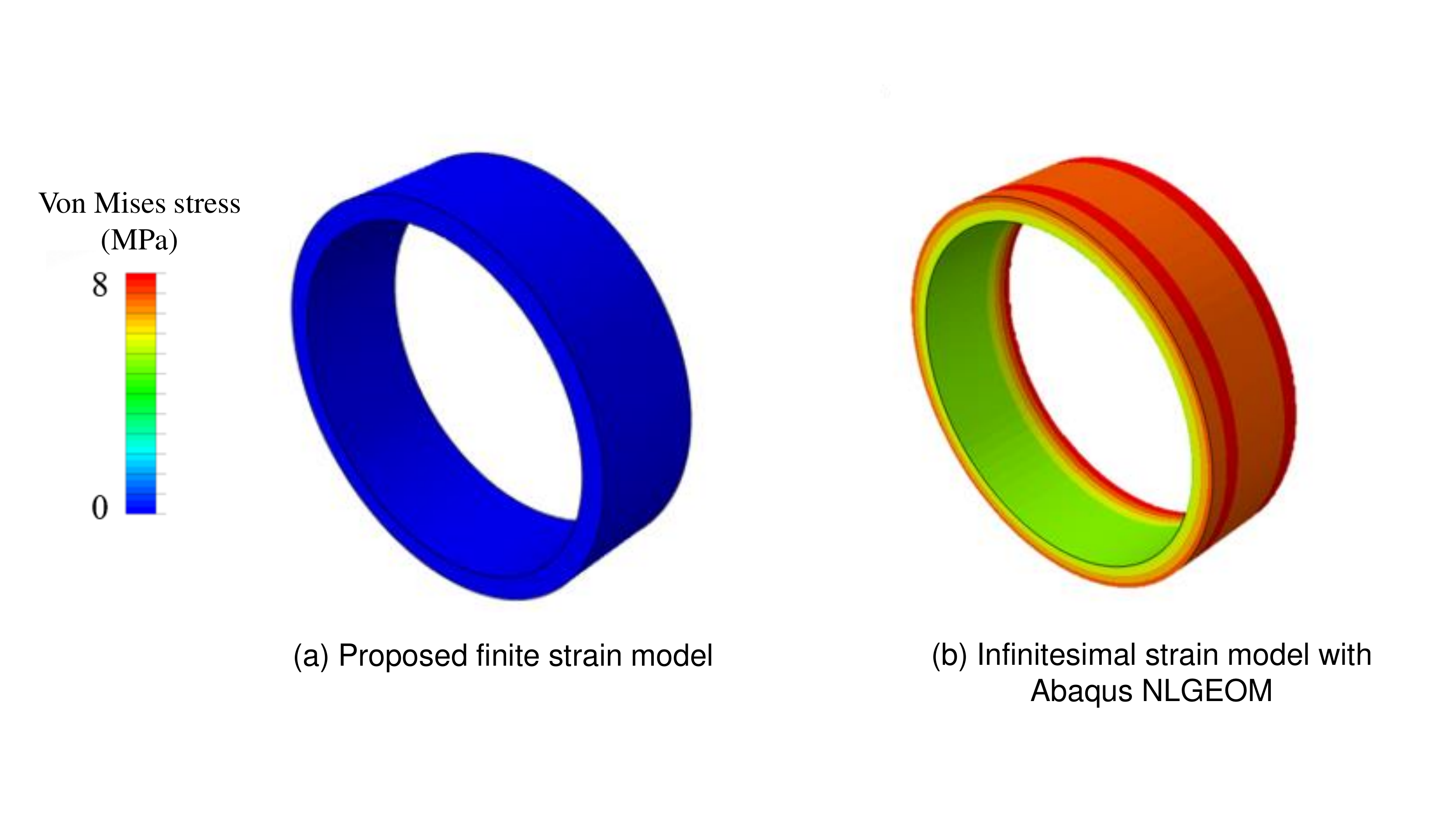}\vspace{-1.3cm}
	\caption{\uppercase{Comparison of artificial residual stress contour predicted by proposed finite strain model and infinitesimal model  with Abaqus NLGEOM option for a SMA torque tube after one loading cycle.}}
	\label{fig:Residual_tube}
\end{figure*}

Pseudoelastic response of the SMA torque tube are examined by proposed finite strain model and its infinitesimal counterpart with Abaqus NLGEOM option. At the end of the first loading cycle, the Von Mises stress are checked in both cases. As it is shown in Fig.\ref{fig:Residual_tube}, the artificial residual stress is almost zero predicted by proposed finite strain model, while the residual stress obtained from infinitesimal model with Abaqus NLGEOM option is around 8 MPa in contrast. The artificial residual stresses from those two cases are summarized in principal components representation at Tab.\ref{table:T_ReStress}.
\begin{table*}[h!]
	\centering\vspace{0.5cm}
	\caption{\uppercase{Artificial residual stress introduced for SMA torque tube after one loading cycle}}
	\label{table:T_ReStress}
	\begin{tabular}{@{}lcccc@{}}
		\toprule
		&{Stress Components}          &{Proposed finite strain model}  & \pbox{30cm}{Infinitesimal strain model with\\ Abaqus NLGEOM} &\\ \midrule
		&Principal Max.(MPa) &1.77e-7        & 4.04           &\\ 
		&Principal Mid.(MPa) &1.05e-7        & -4.19          &\\ 
		&Principal Min.(MPa) &3.22e-8        &-4.32           &\\ 
		\bottomrule
	\end{tabular}
\end{table*}

\begin{figure}[t]
	\vspace{-0.5cm}
	\centering\hspace*{-0.8cm}
	\includegraphics[width=0.5\textwidth]{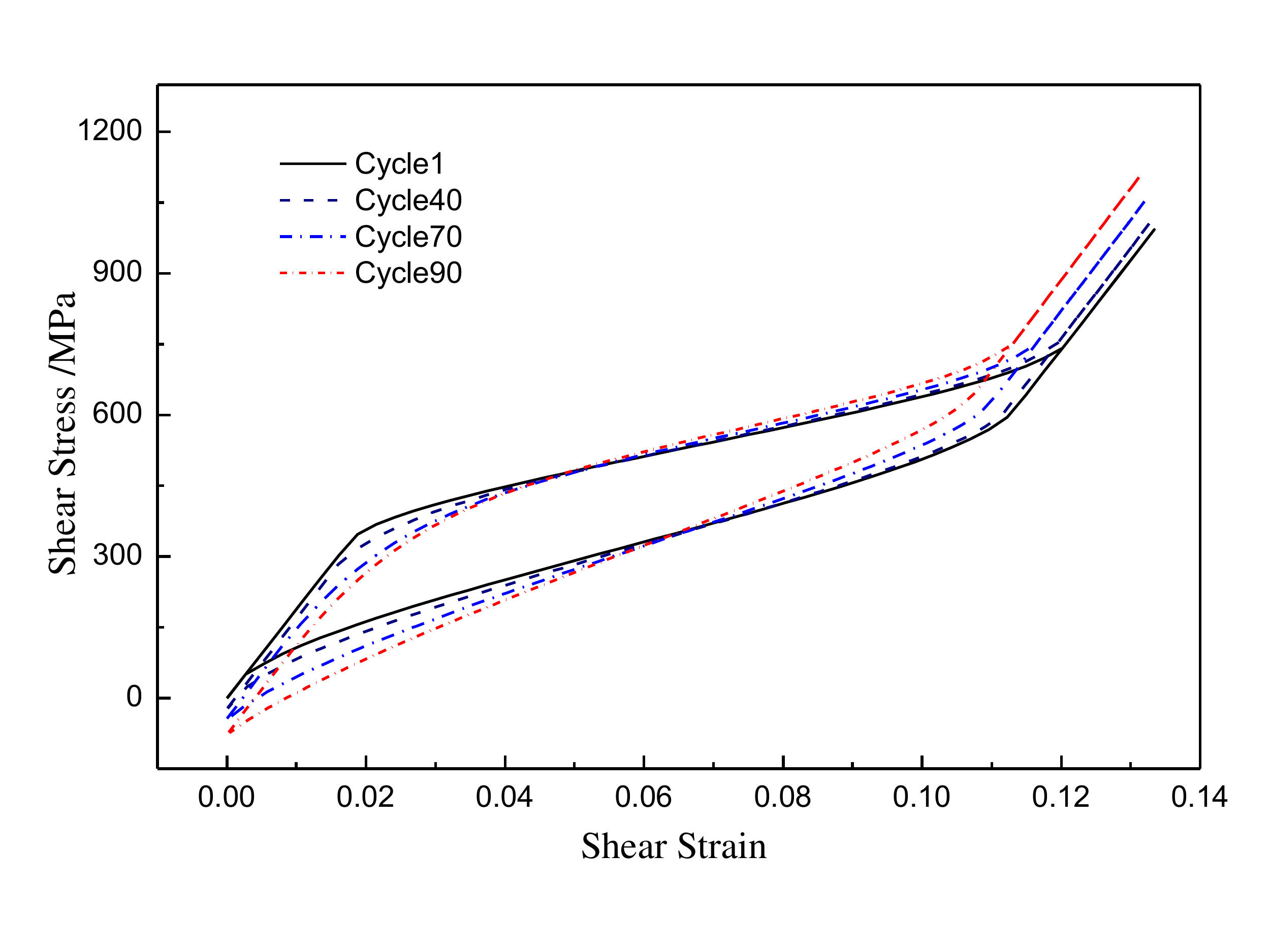}\vspace{-0.7cm}
	\caption{\uppercase{Cyclic pseudoelastic response of a SMA torque tube predicted by infinitesimal model with Abaqus NLGEOM option under 90 cyclic loadings}}
	\label{fig:Tube_ABA_90}
\end{figure}

\begin{figure}[t]
	\vspace{-0.5cm}
	\centering\hspace*{-0.8cm}
	\includegraphics[width=0.5\textwidth]{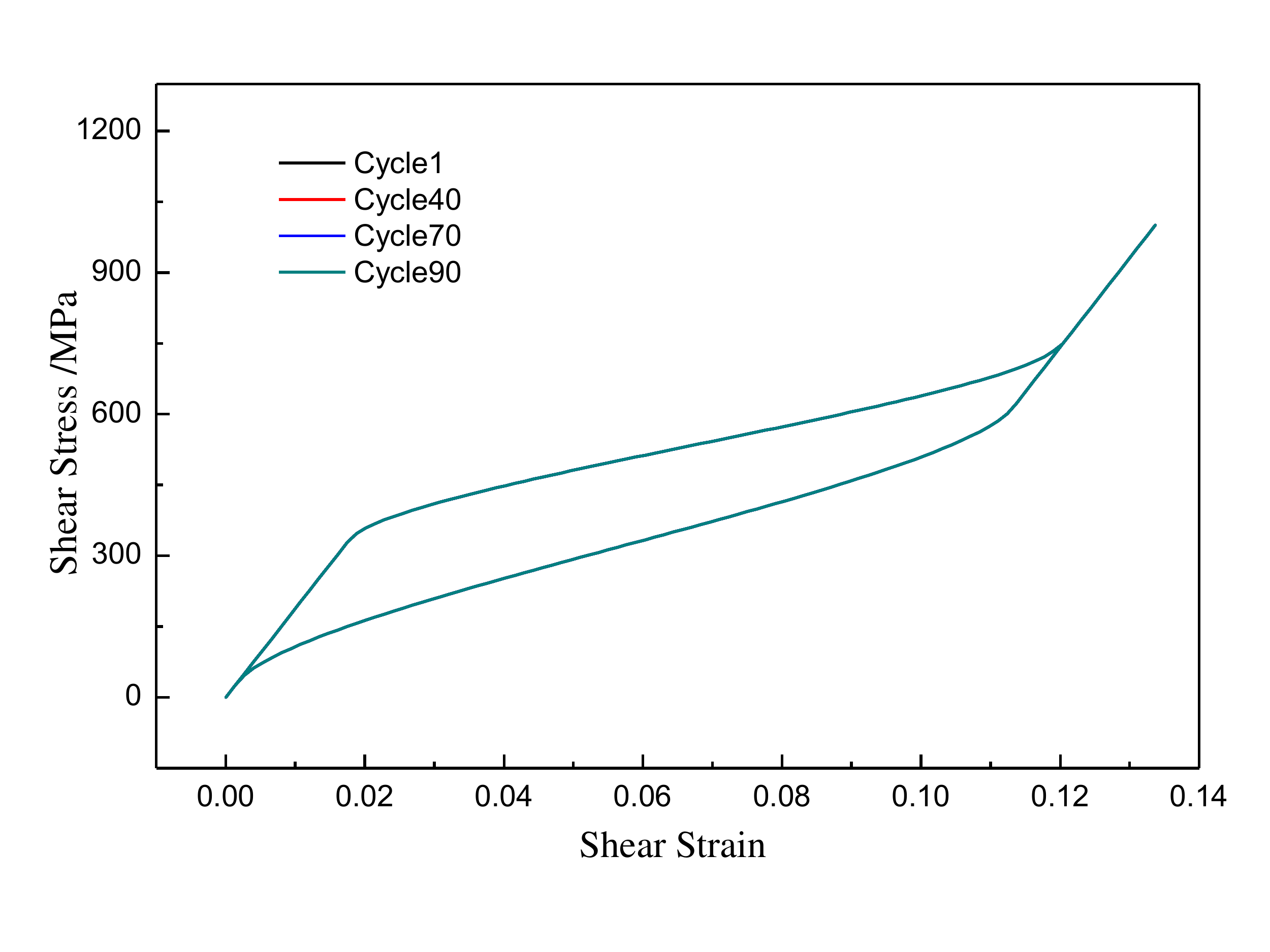}\vspace{-0.7cm}
	\caption{\uppercase{Cyclic Pseudoelastic response of a SMA torque tube predicted by proposed finite strain model under 90 cyclic loadings}}
	\label{fig:Tube_LEI_90}
\end{figure}

Build on the results for one loading cycle, let us now examine whether this accumulation of artificial residual stresses will continue to grow, and how it will affect the material response of SMAs torque tube under cyclic loadings. Referring to Fig.\ref{fig:Tube_ABA_90} for material response predicted by the infinitesimal model using Abaqus NLGEOM option, it can be seen that there is a clear response shifting from initial to final loading cycle due to the accumulation of artificial residual stress. Specifically, the stress levels required to start the forward phase transformation is decreasing, maximum shear stress achieved is increasing and the shear stress at end of each loading cycle is deviating more and more from the initial zero value. In contrast, material response predicted by proposed finite strain model in Fig.\ref{fig:Tube_LEI_90}, the initial material response is nearly overlapping with the material response from subsequent loading cycles.

Based on the results for segment SMA torque tube under cyclic loading, it is demonstrated that, because of the small amount of artificial residual stress introduced by using Abaqus NLGEOM option, the infinitesimal strain based constitutive model for SMAs will predict an shifting, instead of stable, cyclic pseudoelastic response for a torque tube. It is also shown that the proposed finite strain SMA constitutive model, based on logarithmic strain and logarithmic rate, can effectively rule out the effects of artificial residual stress during cyclic loadings. Nevertheless, it is worthy of pointing out for general metallic material within small strain regime during low loading cycles, the amount of artificial residual stresses have quite limited effects on the material response, and it is still a practical and relatively accurate way to utilize the Abaqus NLGEOM option to consider deformation involved with large rotations. At the same time, it is crucial to realized that such undesired artificial stresses need be taken into consideration when the material is undergoing a large number of cyclic loadings, for which a finite strain constitutive model for SMAs based on logarithmic strain and logarithmic rate is indispensible in order to deliver an stable and accurate material response.  
\section*{CONCLUSION}\label{Conc}
Based on the infinitesimal strain based constitutive model for shape memory alloys from Lagoudas and coworkers\cite{boyd1996,lagoudas2012}, a three dimensional phenomenological constitutive model for martensitic transformation in polycrystalline shape memory alloys considering large strains and large rotations is proposed in this work. Numerical simulations considering basic SMA component geometries, such as a bar and a torque tube under stress induced phase transformation, are performed to test the capabilities of the proposed model. For numerical examples of SMA bar, relatively large discrepancies are observed between the responses predicted by the proposed model and its infinitesimal counterpart when strain is no longer considered small. In the numerical simulations of a SMA torque tube with large rotations, it is demonstrated that the difference of predicted material response will be perceivable in moderate strain regime of $5\%$ and continues to grow substantially in large strain regime at $10\%$.  From the results of SMA torque tube under cyclic loading, it has been shown that infinitesimal strain based constitutive model will predict an shifting cyclic pseudoelastic response as a result of the artificial residual stress introduced by using Abaqus NLGEOM option. In comparison, the proposed finite strain SMA constitutive model, based on logarithmic strain and logarithmic rate, can effectively rule out the effects of artificial residual stress during cyclic loadings.

%
%
\section*{Acknowledgments}\label{Ack}
The author would like to acknowledge the financial support provided by the Qatar National Research Fund under the grant number: NPRP 7-032-2-016, and the NASA University Leadership Initiative project under the grant number: NNX17AJ96A. 

\bibliographystyle{asmems4}
\bibliography{myarticle}

\end{document}